%% file: main.tex
\documentclass[a4paper,10pt]{article}
\usepackage{pos}

\usepackage[firstpage]{draftwatermark}
\SetWatermarkFontSize{12pt}
\SetWatermarkColor{black}
\SetWatermarkAngle{0}
\SetWatermarkHorCenter{17.6cm}
\SetWatermarkVerCenter{5.8cm}
\SetWatermarkText{MS-TP-24-16, IFJPAN-IV-2024-10}

\usepackage{cleveref}
\usepackage{wrapfig}
\usepackage{xifthen}
\usepackage{xparse}
\hypersetup{
	colorlinks=true,
	linkcolor=black,
	citecolor=red,
	urlcolor=blue,
	pdftitle={DIS2024: A Markov Chain Monte Carlo determination of Proton PDF uncertainties at NNLO},
	pdfauthor={Risse, P.}
}
\newcommand{\orcid}[1]{\,\href{https://orcid.org/#1}{\includegraphics[width=9pt]{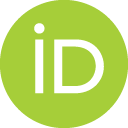}}\,}
\newcommand{\orcidTJ}{0000-0002-1334-7607} 
\newcommand{\orcidKK}{0000-0003-1412-447X} 
\newcommand{\orcidAK}{0000-0002-4090-0084} 
\newcommand{\orcidND}{0000-0003-0962-631X} 
\newcommand{\orcidPR}{0000-0002-8570-5506} 

\newcommand{\apfelxx}{\texttt{APFEL++}}
\newcommand{\applgrid}{\texttt{APPLgrid}}

\newcommand{\fastkernel}{\textsc{FastKernel}}
\newcommand{\anti}[1]{\ensuremath{\bar{#1}}}
\newcommand{\fitparam}[1]{\texttt{#1}}
\newcommand{\dvpfour}{$p_4$}

\NewDocumentCommand{\observable}{O{} O{} O{}}{\ifthenelse{\isempty{#3}}
	{\ensuremath{\mathcal{O}_{#1}^{#2}}}
	{\ensuremath{\mathcal{O}_{#1}^{#2}\left(#3\right)}}}
\NewDocumentCommand{\observablelow}{O{} O{} O{}}{\ifthenelse{\isempty{#1}}
	{\observable[-][#2][#3]}
	{\observable[-,#1][#2][#3]}}
\NewDocumentCommand{\observableup}{O{} O{} O{}}{\ifthenelse{\isempty{#1}}
	{\observable[+][#2][#3]}
	{\observable[+,#1][#2][#3]}}

\title{A Markov Chain Monte Carlo determination of Proton PDF uncertainties at NNLO}
\ShortTitle{A MCMC determination of Proton PDF uncertainties at NNLO}

\author*[a]{Peter Risse\orcid{\orcidPR}}
\author[b]{Nasim Derakhshanian\orcid{\orcidND}}
\author[a]{Tomas Je\v{z}o\orcid{\orcidTJ}}
\author[a]{Karol Kova\v{r}\'{i}k\orcid{\orcidKK}}
\author[b]{Aleksander Kusina\orcid{\orcidAK}}

\affiliation[a]{Institut f{\"u}r Theoretische Physik, Universit{\"a}t M{\"u}nster,\\
	Wilhelm-Klemm-Stra{\ss}e 9, D-48149 M{\"u}nster, Germany}

\affiliation[b]{Institute of Nuclear Physics Polish Academy of Sciences,\\
	PL-31342 Krakow, Poland}

\emailAdd{risse.p@uni-muenster.de}

\abstract{The current scientific standard in PDF uncertainty estimation relies either on repeated fits over artificially generated data to arrive at Monte Carlo samples of best fits or on the Hessian method, which uses a quadratic expansion of the figure of merit, the 
	$\chi^2$-function. Markov Chain Monte Carlo methods allow one to access the uncertainties of PDFs without making use of quadratic approximations in a statistically sound procedure while at the same time preserving the correspondence between the sample and $\chi^2$-value.
	Rooted in Bayesian statistics the $\chi^2$-function is repeatedly sampled to obtain a set of PDFs that serves as a representation of the statistical distribution of the PDFs in their function space. After removing the dependence between the samples (the so-called autocorrelation) the set can be used to propagate the uncertainties to physical observables.
	The final result is an independent procedure to obtain PDF uncertainties that can be confronted by the state-of-the-art in order to ultimately arrive at a better understanding of the proton's structure.}

\FullConference{31st International Workshop on Deep Inelastic Scattering (DIS2024)\\
	8–12 April 2024\\
	Grenoble, France\\}

\begin{document}
	\maketitle

	\section{Introduction}
	
	In recent years proton PDF extractions have become more and more precise by including newer and more complete experimental data sets, utilizing more experimental observables, increasing the theoretical accuracy from NNLO to approximate N3LO and including further methodical advancements. Recently, also the estimation of the error PDFs has re-gained interest \cite{Kovarik:2019xvh,Hunt-Smith:2022ugn}, including the proposal of using advanced statistical tools of uncertainty estimation: Markov Chain Monte Carlo (MCMC). So far this method has only been used in toy models \cite{Hunt-Smith:2023ccp,Capel:2024qkm} or in extractions using only DIS data from HERA run I and II \cite{Gbedo:2017eyp}, because the analysis is much more involved computationally. In this talk we present a proton PDF uncertainty estimation from MCMC with a realistic set of data and compare the results with the state-of-the-art for global analyses.
	
	\section{Experimental data and theoretical setup}
	
\begin{wrapfigure}{r}{0.6\linewidth}
		\centering
		\includegraphics[width=\linewidth]{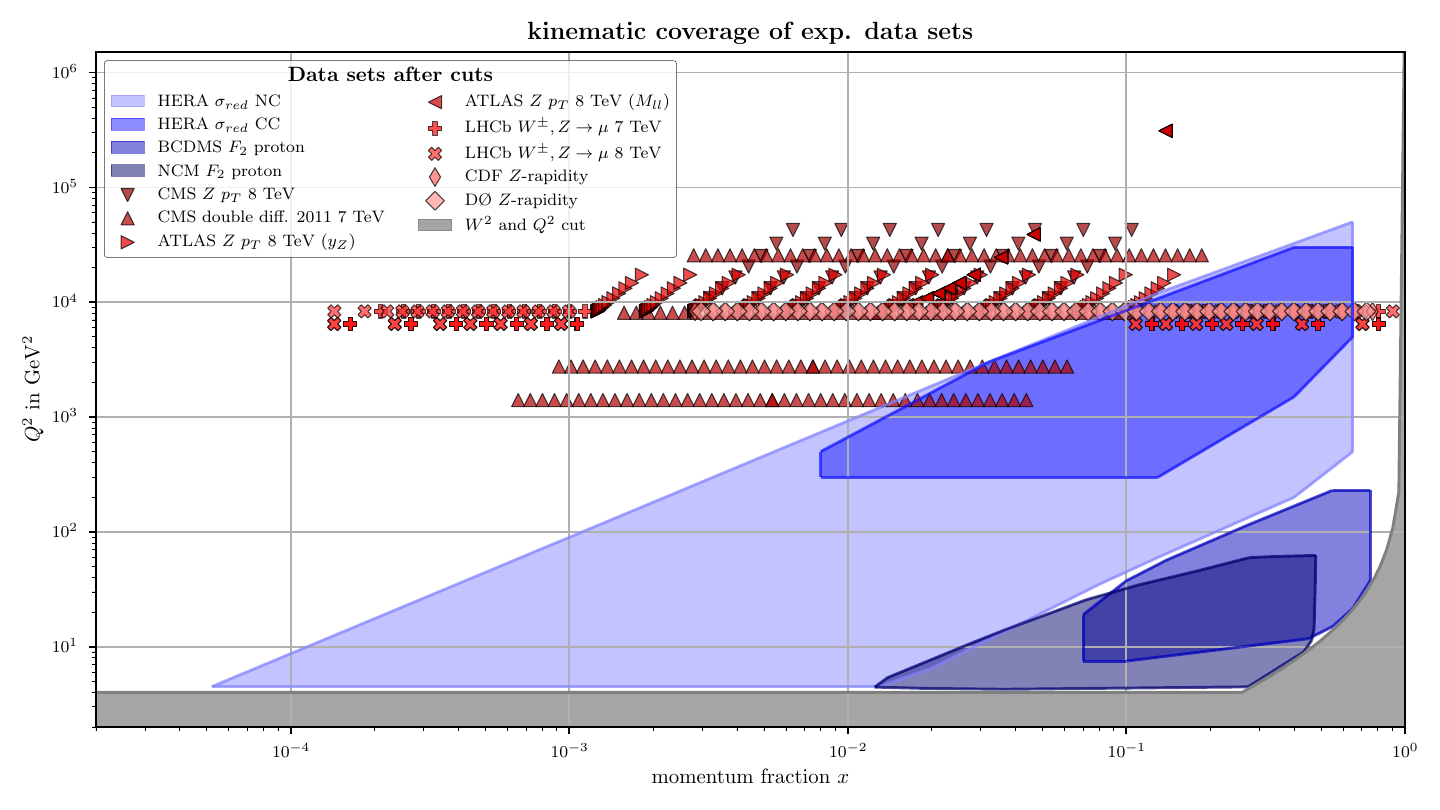}
		\caption{The kinematic coverage of the experimental data sets (see \cref{tab:exp_data}) used for the extraction. DIS data sets are indicated by blue patches, DY data sets by red by individual symbols and cut regions in gray. We used leading order approximations to estimate the $(x,Q^2)$-point.}
		\label{fig:kinematic_coverage}
	\end{wrapfigure}

	The goal of the analysis is to perform a realistic proton PDF extraction, whilst keeping the computational effort at a reasonable level. As a compromise we only consider a reduced selection of experimental data sets compared to a global PDF analysis, because we exclusively rely on theoretical predictions in the form of fast-convolution grids. This allows for an extremely fast recalculation of theoretical predictions and ultimately makes the statistical investigation with Markov Chains possible.
	The complete list of data sets is given in \cref{tab:exp_data} grouped by observable; the kinematic coverage is given in \cref{fig:kinematic_coverage}. In all cases we use NNLO theoretical accuracy and take correlated uncertainties into account wherever available.
	
	In the following we give a brief description of the considerations behind the selection. Finally we introduce our PDF parametrization.
	
	\paragraph{Deep inelastic scattering}
	The DIS data come as measurements of the neutral current $F_2$ structure function (BCDMS, NMC) and as the reduced cross section (combined data set of H1 and ZEUS) for neutral and charged current. The theoretical predictions are obtained in the aSACOT-$\chi$ mass-scheme \cite{Stavreva:2012bs} that was recently implemented in the numerical library \apfelxx{} \cite{Risse:2023rxd,Risse:2024hqme}. In this library the numerical predictions are pre-calculated in tables that only have to be interpolated by a PDF set at runtime. This yields a massive speed advantage compared to naive implementations. We employ kinematic cuts of 
	\begin{equation}
		W^2 \geq 12.25\,\text{GeV}^2 \quad \text{and} \quad Q^2 \geq 4 \,\text{GeV}^2\,.
	\end{equation}
	The kinematic coverage of the data is given by the blue patches in \cref{fig:kinematic_coverage}.
	
	\paragraph{Drell-Yan}
	In this setup we use the NLO \applgrid{}-tables published in Ref.\,\cite{NNPDFAPPLgrids} by the NNPDF collaboration and translate these to \fastkernel{}-tables \cite{Bertone:2016lga} in order to increase the evaluation speed by $\mathcal{O}(100)$ over the \applgrid{}-tables. NNLO accuracy is achieved by using $K$-factors, which have been published by NNPDF in Refs.\,\cite{NNPDFcode,NNPDF:2017mvq}. In order to stay consistent, we align our cuts with the NNPDF analysis, see Ref.\,\cite[table 2.4]{NNPDF:2017mvq}. The data sets are represented by the red symbols in \cref{fig:kinematic_coverage}.
	
	\begin{table}[t]
		\centering
		\footnotesize
		\input{tab_fit_result}
		\caption{The experimental data sets considered alongside with the $\chi^2/\rm{DATA}$ value of the \textit{best-fit} sample. In total the analysis included 1984 data points (after kinematic cuts), which the MCMC samples are able to describe up to a $\chi^2$-value of 1.2 per degree of freedom.}
		\label{tab:exp_data}
		\normalsize
	\end{table}
	
	\paragraph{PDF parametrization}
	Motivated by the proton PDF extractions in the CJ family \cite{Accardi:2016qay,Accardi:2023gyr}, we use the parametric form 
	\begin{equation}
		xf(x,Q_0) = p_0x^{p_1}(1-x)^{p_2}\left(1+p_3\sqrt{x}+p_4x\right)
	\end{equation}
	for the flavor combinations $u_v,d_v,\anti{d}+\anti{u},g$ and $s+\anti{s}$, where $p_{0\dots 4}$ are the fit-able parameters. The parametrization scale $Q_0$ is placed at the charm threshold $m_c=1.3\,\text{GeV}$. In $u_v,d_v$ and $\anti{d}+\anti{u}$ we fix $p_0$ through sum rules and set $p_3$ to $\{0,-3.503,0\}$ respectively. For the gluon distribution we open all five parameters and for the strange-combination only $p_0$, whilst keeping $p_{1\dots4}$ fixed at $\{-0.20775,0,0,14.606\}$. Lastly we add $p_5x^{p_6}xu_v(x,Q_0)$ to the down-valence distribution, where we set $p_5= 0.0036$ and $p_6=2$. In total we fit 15 parameters, whose vector we denote in the following by $\fitparam{p}$.

	\section{Markov Chain Monte Carlo setup}
	
	Rooted in Bayesian statistics, the probability density of finding the 15 PDF parameters $\fitparam{p}$ given the experimental data $\{D_i\}$ can be written as
	\begin{equation}
		\pi(\fitparam{p}|D) = \frac{1}{\mathcal{N}}p(\fitparam{p})\exp\left(-\frac{1}{2}\chi^2(D,T(\fitparam{p}))\right)\,.
		\label{eq:parameter_prob_density}
	\end{equation}
	Here $\chi^2$ is the usual $\chi^2$-function (e.g. \cite{Kovarik:2019xvh}) taking correlated and normalization uncertainties into account, $T(\fitparam{p})$ are the theoretical predictions calculated from the PDF parameters, $\mathcal{N}$ is an irrelevant normalization constant and $p(\fitparam{p})$ is the prior distribution for the PDF parameters, which we define later.
	
	The goal of the MCMC analysis is to find a set of parameter samples $\fitparam{p}_t$ (with $t=1\dots N$), which approximates the probability distribution $\pi(\fitparam{p}|D)$, i.e.\;the samples approximate expectation values of any observable $\mathcal{O}(\fitparam{p})$:
	\begin{equation}
		\langle \mathcal{O}(\fitparam{p}) \rangle_\pi = \int \mathrm{d}\fitparam{p}\,\mathcal{O}(\fitparam{p})\pi(\fitparam{p}) \approx \frac{1}{N} \sum_{t}^{N}\mathcal{O}(\fitparam{p}_t) = \frac{1}{N}\sum_{t}^{N}\mathcal{O}_t\,.
		\label{eq:exp_value_observable}
	\end{equation}
	
	The MCMC samples are generated in a procedural algorithm, where every new sample $\fitparam{p}_{t+1}$ is first proposed from the current sample by the adaptive Metropolis-Hastings algorithm \cite{Haario:2001} and then accepted/rejected by the Metropolis-Hastings \cite{Metropolis:1953,Hastings1970} acceptance probability $a(\fitparam{p}_{t},\fitparam{p}_{t+1})$. Here, $a(\fitparam{p}_{t},\fitparam{p}_{t+1})$ includes information from \cref{eq:parameter_prob_density} such that the density of the samples follows the density given by the experimental measurements. However, each newly proposed sample requires a re-evaluation of the $\chi^2$-function making the procedure computationally expensive. If $\fitparam{p}_{t+1}$ is accepted, it gets appended to the list, otherwise the current sample is repeated. Since $\fitparam{p}_{t+1}$ was proposed based on $\fitparam{p}_{t}$, there exists a dependence between the samples, which is called autocorrelation (see e.g.\;Ref.\,\cite{HandbookMCMC:2011}) and can be intuitively understood as a reduced gain of information compared to the information gained from an independently generated (i.e. uncorrelated) sample. The starting point can be chosen (or generated) freely. In the beginning the chain will drift towards the region of highest probability. This is so-called thermalization time has to be removed to avoid bias.
	
	In the following we briefly discuss the priors for the parameters, the global settings for generating the chain and the purification, where we remove the thermalization region and autocorrelation to ultimately arrive at a reduced set of samples that can be used in the uncertainty estimation.
	
	\paragraph{Priors}
	We set the priors of the parameters to a constant value (which is absorbed by the normalization constant), except for the prior for $p_4$ of the down-valence distribution. This parameter value is unconstrained from above by the experimental data due to a flaw in the parametrization, where the PDF becomes effectively independent of \dvpfour{} if it becomes too large. Instead of fixing its value, we use a uniform prior with the bounds $\left[-10^3,10^4\right]$\footnote{Only the upper limit is relevant, as the parameter is constraint by the data from below.}, which sets the acceptance probability to zero, if it is proposed outside of the limits. The prior is constant if \dvpfour{} is proposed inside the limits and thus absorbed by the normalization keeping the correspondence between the sample and the $\chi^2$-value intact.
	
	\paragraph{Generating the samples}
	We generate 36 independent chains, each consisting of 479,000 samples yielding 17,244,000 in total after 14 days of computing time (on 36 cores in parallel). The starting point for each chain was found by first running a minimization algorithm to find the region of highest probability and then individually perturbing them in a random fashion from the minimum to keep the chains independent. The proposal algorithm was reset at the 20,000th and 40,000th step to boost convergence.
	
	\paragraph{Purification}
	The thermalization is roughly finished after the 120,000th iteration. To be conservative, we choose to remove the data before the 140,000 sample. As the chain exhibits strong autocorrelation, we thin the chain, i.e. instead of using every sample we only consider every $\eta$-th sample. This does not only reduce the chain and therefore the computational costs of \cref{eq:exp_value_observable} greatly, but also simplifies the interpretation of each sample individually. It is to be noted that thinning reduces the statistics \cite{Geyer:1992pmcmc,Link:2012thinning} (i.e.\;our results are less precise after thinning), but we still end up with a sufficiently large sample. 
	
	The autocorrelation is estimated by the $\Gamma$-method \cite{Wolff2004} and after applying a thinning factor of $\eta=3000$ (for each chain individually) we arrive at the estimate that the next independent sample is on average found after $2\tau_{int} = 1.14 \pm 0.07$ steps, very close to the optimal value of one. Applying a larger thinning factor does not yield improved results. The final number of samples is $N=4068$, which we consider as approximately uncorrelated and free from thermalization bias.
	
	\section{Final PDF uncertainty estimation}
	
	The uncertainty estimation on observables or the PDFs themselves based on the samples can be carried out in several ways. Usually the task is to find a central value $\observable[*]$ along with the confidence interval $\left[\observablelow,\observableup\right]$ corresponding to some probability $p$, often $p=90\%$. One of the simplest symmetric estimations is based on the moments of the observable: $\langle\mathcal{O}\rangle \pm z_p\sqrt{\langle\mathcal{O}^2-\langle\mathcal{O}\rangle^2\rangle}$, where $z_p$ is the quantile for $p$. Following Ref.\,\cite{Gbedo:2017eyp}, an asymmetric estimation can be defined by setting the central value to the best fit sample and then performing a quantile estimation on the upper and lower value of $\observable$. 
	
\begin{wrapfigure}{r}{0.5\linewidth}
		\centering
		\includegraphics[width=\linewidth]{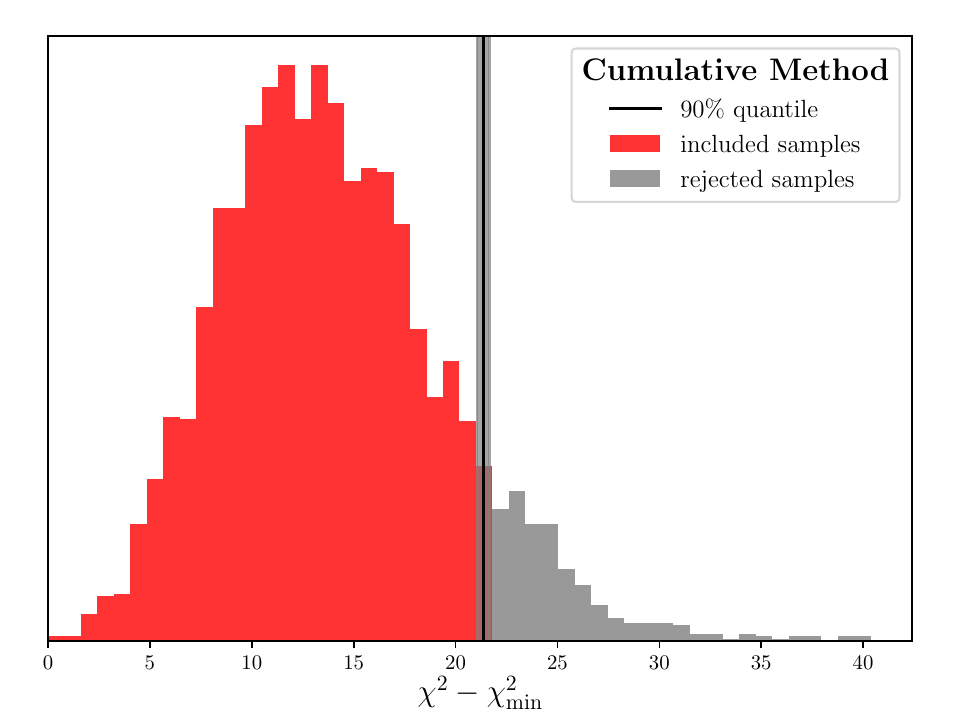}
		\caption{$90\%$-quantile estimation of the distribution of the $\chi^2$-values.}
		\label{fig:chi2_quantile_estimation}
	\end{wrapfigure}

	Here we follow the definition of Ref.\,\cite{Putze:2008ps}, which we call the ``Cumulative $\chi^2$''-method: The central value is set to the best fit sample (i.e. the sample with the minimal $\chi^2$-value). Then we perform a quantile estimation of the distribution of the $\chi^2$-values as depicted in \cref{fig:chi2_quantile_estimation}. The lower/upper bound on the observable is defined as the minimal/maximal value found within the quantile.
	With our samples we find $\chi^2_{\max}=22$ for the $90\%$ quantile, which is in agreement with a $\chi^2$-distribution with 15 degrees of freedom. Intuitively the confidence interval of this method can be understood as the ``maximum reach'' an observable can have, whilst still being in the $90\%$-quantile of describing the experimental data.
	
\begin{figure}[h]
		\centering
		\includegraphics[width=0.5\linewidth]{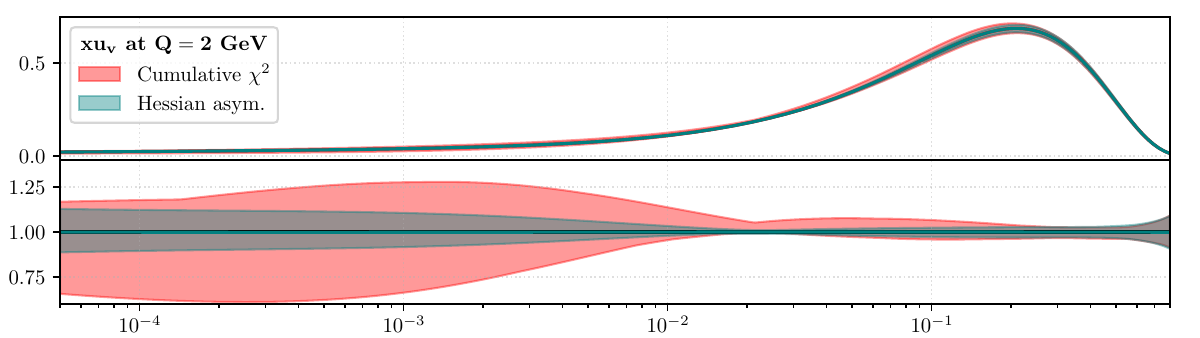}%
		\includegraphics[width=0.5\linewidth]{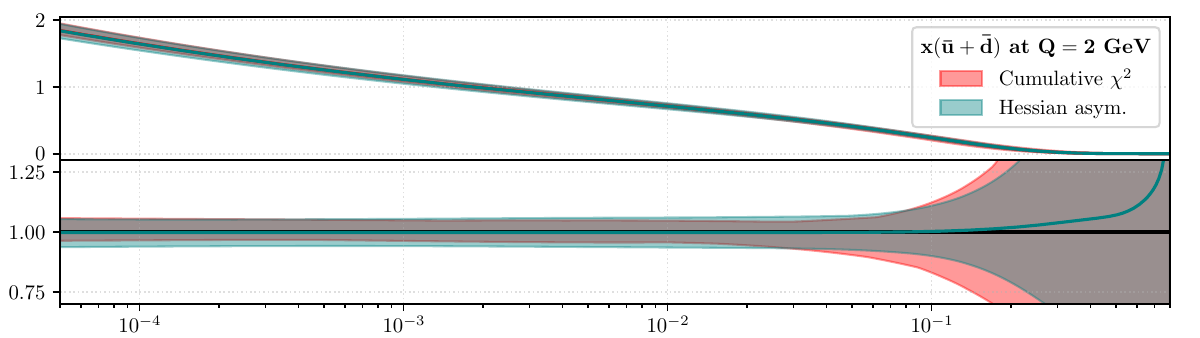}
		\includegraphics[width=0.5\linewidth]{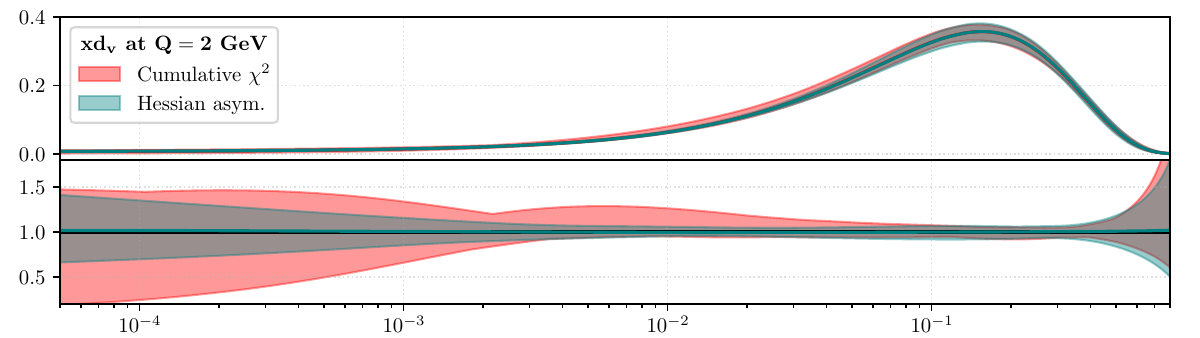}%
		\includegraphics[width=0.5\linewidth]{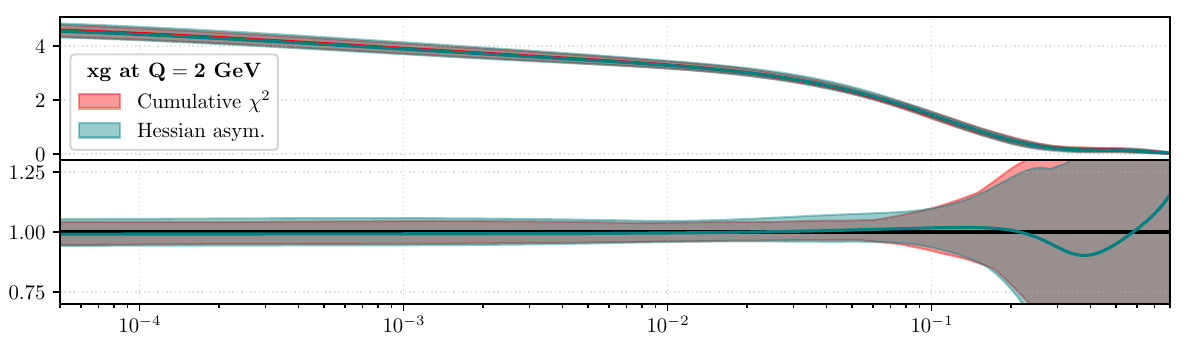}
		\caption{Comparison of PDF error estimations for the $u_v,d_v,\anti{d}+\anti{u}$ and gluon distribution using the Cumulative $\chi^2$ (red) and Hessian (blue) method. The upper panels show the absolute PDFs and the bottom shows the ratio to the central value of the Cumulative $\chi^2$ method.}
		\label{fig:PDF_uncertainty_comparison}
	\end{figure}

	Finally, we compare the MCMC uncertainty estimation with the Hessian method \cite{Pumplin:2001ct}. Thus we employ a standard minimization algorithm and calculate the asymmetric error PDFs. For this purpose we need to set a tolerance for the error PDFs, which the Hessian method does not provide. 
	Therefore we resort to the MCMC analysis and use the quantile estimation of the distribution of the $\chi^2$-values and set $\Delta\chi^2_{\rm Hessian} = \chi^2_{\max}$. 
	In \cref{fig:PDF_uncertainty_comparison} we show the error PDFs for both methods for the $u_v,d_v$ (left) and $\anti{u}+\anti{d}, g$ (right) distributions. 
	From the ratio plots (lower panels) we can see that the error estimations agree on the right figures, whilst the cumulative method gives larger uncertainties on the left. 
	This agrees with the marginal distributions of the parameters: The parameters corresponding to $\anti{u}+\anti{d}, g$ follow Gaussian distributions closely and can therefore be captured by the Hessian method.
	This no longer holds for the parameters of $u_v, d_v$.
	Here the marginal distributions differ from Gaussian significantly and are therefore not captured well by the Hessian method. 
	Correspondingly the uncertainties do not agree, with the Hessian ones being markedly smaller, in some regions more than a factor of two.
	Even though the tolerance is in principle a free parameter of the Hessian method, increasing its value such that the uncertainty bands from the two different methods agree for valence distributions, would lead to overestimated uncertainty in the anti-quark and gluon distributions.
	In any case: Without the MCMC analysis this issue would have been very hard to catch as the marginal parameter distributions are not available in the Hessian method. 
	Instead, one- or two-dimensional parameter scans are performed, which keep the remaining parameter values fixed and are therefore difficult to interpret correctly.
	
	We conclude that although a Markov Chain Monte Carlo analysis is computationally intensive, it yields the benefits of a more sophisticated statistical analysis tool: We obtain an independent procedure to estimate PDF uncertainties without approximations, which agrees with the Hessian method in regions, where its approximations hold and brings insights in regions, where the approximations break. Furthermore, this analysis can be used to estimate the tolerance.
	
	\clearpage
	\section*{Acknowledgements}
	
	Calculations (or parts of them) for this publication were performed on the HPC cluster PALMA II of the University of Münster, subsidised by the DFG (INST 211/667-1). 
	P.R., T.J. and K.K. acknowledge support of the DFG through the Research Training Group GRK 2149. 
	A.K. and N.D. are grateful for the support of Narodowe Centrum Nauki under grant no. 2019/34/E/ST2/00186.
	
	\bibliographystyle{JHEP}
	\bibliography{bib}
	
	
\end{document}

%% file: tab_fit_result.tex
\begin{tabular}{lccc|lccc}
	\textsc{Data Set}  & \textsc{Ref.} & \textsc{Points} & $\chi^2$\textsc{/data} &\textsc{Data Set}  & \textsc{Ref.} & \textsc{Points} & $\chi^2$\textsc{/data}\\
	\hline
	\textbf{DIS} & & & &\textbf{DY} & & &\\
	HERA $\sigma_{red}$ neutral current & \cite{H1:2015ubc} & 1039&1.26 & CDF $Z$-rapidity & \cite{CDF:2010vek} & 28 &1.10\\
	HERA $\sigma_{red}$ charged current & \cite{H1:2015ubc} & 81&1.08 & D{\O} $Z$-rapidity & \cite{D0:2007djv} & 28&0.60\\
	BCDMS $F_2$ proton & \cite{BCDMS:1989qop} & 339&1.09 & ATLAS $Z$ $p_T$ 8 TeV ($M_{ll}$) & \cite{ATLAS:2015iiu} & 44&1.06\\
	NCM $F_2$ proton & \cite{NewMuon:1996fwh} & 201&1.54 & ATLAS $Z$ $p_T$ 8 TeV ($y_Z$) & \cite{ATLAS:2015iiu} & 48&0.65\\
         & & & & CMS $Z$ $p_T$ 8 TeV & \cite{CMS:2015hyl} & 28&0.46\\
         & & & & CMS double diff. 2011 7 TeV & \cite{CMS:2013zfg} & 88&1.02\\
         & & & & LHCb $W^\pm,Z\rightarrow \mu$ 7 TeV & \cite{LHCb:2015okr} & 29&1.07\\
         & & & & LHCb $W^\pm,Z\rightarrow \mu$ 8 TeV & \cite{LHCb:2015mad} & 31&1.18\\[4pt]
         
	DIS total & & 1660&1.25 & DY total & & 324&0.91\\
	\hline
	\textbf{Total} & & 1984 & \multicolumn{5}{l}{\enspace\,\,\textbf{1.20} (per dof)} 
\end{tabular}

%% file: main.bbl
\providecommand{\href}[2]{#2}\begingroup\raggedright\begin{thebibliography}{10}

\bibitem{Kovarik:2019xvh}
K.~Kova\v{r}\'{i}k, P.M.~Nadolsky and D.E.~Soper, \emph{{Hadronic structure in
  high-energy collisions}},
  \href{https://doi.org/10.1103/RevModPhys.92.045003}{\emph{Rev. Mod. Phys.}
  {\bfseries 92} (2020) 045003}
  [\href{https://arxiv.org/abs/1905.06957}{{\ttfamily 1905.06957}}].

\bibitem{Hunt-Smith:2022ugn}
N.T.~Hunt-Smith, A.~Accardi, W.~Melnitchouk, N.~Sato, A.W.~Thomas and
  M.J.~White, \emph{{Determination of uncertainties in parton densities}},
  \href{https://doi.org/10.1103/PhysRevD.106.036003}{\emph{Phys. Rev. D}
  {\bfseries 106} (2022) 036003}
  [\href{https://arxiv.org/abs/2206.10782}{{\ttfamily 2206.10782}}].

\bibitem{Hunt-Smith:2023ccp}
N.T.~Hunt-Smith, W.~Melnitchouk, F.~Ringer, N.~Sato, A.W.~Thomas and
  M.J.~White, \emph{{Accelerating Markov Chain Monte Carlo sampling with
  diffusion models}},
  \href{https://doi.org/10.1016/j.cpc.2023.109059}{\emph{Comput. Phys. Commun.}
  {\bfseries 296} (2024) 109059}
  [\href{https://arxiv.org/abs/2309.01454}{{\ttfamily 2309.01454}}].

\bibitem{Capel:2024qkm}
F.~Capel, R.~Aggarwal, M.~Botje, A.~Caldwell, O.~Schulz and A.~Verbytskyi,
  \emph{{PartonDensity.jl: a novel parton density determination code}},
  \href{https://arxiv.org/abs/2401.17729}{{\ttfamily 2401.17729}}.

\bibitem{Gbedo:2017eyp}
Y.G.~Gbedo and M.~Mangin-Brinet, \emph{{Markov chain Monte Carlo techniques
  applied to parton distribution functions determination: Proof of concept}},
  \href{https://doi.org/10.1103/PhysRevD.96.014015}{\emph{Phys. Rev. D}
  {\bfseries 96} (2017) 014015}
  [\href{https://arxiv.org/abs/1701.07678}{{\ttfamily 1701.07678}}].

\bibitem{Stavreva:2012bs}
T.~Stavreva, F.I.~Olness, I.~Schienbein, T.~Jezo, A.~Kusina, K.~Kovarik et~al.,
  \emph{{Heavy Quark Production in the ACOT Scheme at NNLO and N3LO}},
  \href{https://doi.org/10.1103/PhysRevD.85.114014}{\emph{Phys. Rev. D}
  {\bfseries 85} (2012) 114014}
  [\href{https://arxiv.org/abs/1203.0282}{{\ttfamily 1203.0282}}].

\bibitem{Risse:2023rxd}
P.~Risse, V.~Bertone, T.~Je\v{z}o, M.~Klasen, K.~Kova\v{r}\'\i{}k, F.I.~Olness
  et~al., \emph{{Fast evaluation of heavy-quark contributions to DIS in
  APFEL++}},  in \emph{{30th International Workshop on Deep-Inelastic
  Scattering and Related Subjects}}, 7, 2023
  [\href{https://arxiv.org/abs/2307.08269}{{\ttfamily 2307.08269}}].

\bibitem{Risse:2024hqme}
P.~Risse, V.~Bertone, T.~Jezo, K.~Kovarik, F.I.~Kusina, A.~Olness and
  I.~Schienbein, ``{Heavy Quark mass effects in charged current Deep-Inelastic
  Scattering at NNLO in the ACOT scheme}.'' (in preparation).

\bibitem{NNPDFAPPLgrids}
{\scshape NNPDF} collaboration, ``applgrids.''
  \url{https://github.com/NNPDF/applgrids}, 2017.

\bibitem{Bertone:2016lga}
V.~Bertone, S.~Carrazza and N.P.~Hartland, \emph{{APFELgrid: a high performance
  tool for parton density determinations}},
  \href{https://doi.org/10.1016/j.cpc.2016.10.006}{\emph{Comput. Phys. Commun.}
  {\bfseries 212} (2017) 205}
  [\href{https://arxiv.org/abs/1605.02070}{{\ttfamily 1605.02070}}].

\bibitem{NNPDFcode}
{\scshape NNPDF} collaboration, ``nnpdf.''
  \url{https://github.com/NNPDF/nnpdf}, 2017.

\bibitem{NNPDF:2017mvq}
{\scshape NNPDF} collaboration, \emph{{Parton distributions from high-precision
  collider data}},
  \href{https://doi.org/10.1140/epjc/s10052-017-5199-5}{\emph{Eur. Phys. J. C}
  {\bfseries 77} (2017) 663}
  [\href{https://arxiv.org/abs/1706.00428}{{\ttfamily 1706.00428}}].

\bibitem{H1:2015ubc}
{\scshape H1, ZEUS} collaboration, \emph{{Combination of measurements of
  inclusive deep inelastic ${e^{\pm }p}$ scattering cross sections and QCD
  analysis of HERA data}},
  \href{https://doi.org/10.1140/epjc/s10052-015-3710-4}{\emph{Eur. Phys. J. C}
  {\bfseries 75} (2015) 580}
  [\href{https://arxiv.org/abs/1506.06042}{{\ttfamily 1506.06042}}].

\bibitem{CDF:2010vek}
{\scshape CDF} collaboration, \emph{{Measurement of $d\sigma/dy$ of Drell-Yan
  $e^+e^-$ pairs in the $Z$ Mass Region from $p\bar{p}$ Collisions at
  $\sqrt{s}=1.96$ TeV}},
  \href{https://doi.org/10.1016/j.physletb.2010.06.043}{\emph{Phys. Lett. B}
  {\bfseries 692} (2010) 232}
  [\href{https://arxiv.org/abs/0908.3914}{{\ttfamily 0908.3914}}].

\bibitem{D0:2007djv}
{\scshape D0} collaboration, \emph{{Measurement of the Shape of the Boson
  Rapidity Distribution for $p \bar{p} \to Z / \gamma^* \to e^{+} e^{-} + X$
  Events Produced at $\sqrt{s}$ of 1.96-TeV}},
  \href{https://doi.org/10.1103/PhysRevD.76.012003}{\emph{Phys. Rev. D}
  {\bfseries 76} (2007) 012003}
  [\href{https://arxiv.org/abs/hep-ex/0702025}{{\ttfamily hep-ex/0702025}}].

\bibitem{BCDMS:1989qop}
{\scshape BCDMS} collaboration, \emph{{A High Statistics Measurement of the
  Proton Structure Functions F(2) (x, Q**2) and R from Deep Inelastic Muon
  Scattering at High Q**2}},
  \href{https://doi.org/10.1016/0370-2693(89)91637-7}{\emph{Phys. Lett. B}
  {\bfseries 223} (1989) 485}.

\bibitem{ATLAS:2015iiu}
{\scshape ATLAS} collaboration, \emph{{Measurement of the transverse momentum
  and $\phi ^*_{\eta }$ distributions of Drell\textendash{}Yan lepton pairs in
  proton\textendash{}proton collisions at $\sqrt{s}=8$ TeV with the ATLAS
  detector}}, \href{https://doi.org/10.1140/epjc/s10052-016-4070-4}{\emph{Eur.
  Phys. J. C} {\bfseries 76} (2016) 291}
  [\href{https://arxiv.org/abs/1512.02192}{{\ttfamily 1512.02192}}].

\bibitem{NewMuon:1996fwh}
{\scshape New Muon} collaboration, \emph{{Measurement of the proton and
  deuteron structure functions, F2(p) and F2(d), and of the ratio sigma-L /
  sigma-T}}, \href{https://doi.org/10.1016/S0550-3213(96)00538-X}{\emph{Nucl.
  Phys. B} {\bfseries 483} (1997) 3}
  [\href{https://arxiv.org/abs/hep-ph/9610231}{{\ttfamily hep-ph/9610231}}].

\bibitem{CMS:2015hyl}
{\scshape CMS} collaboration, \emph{{Measurement of the Z boson differential
  cross section in transverse momentum and rapidity in
  proton\textendash{}proton collisions at 8 TeV}},
  \href{https://doi.org/10.1016/j.physletb.2015.07.065}{\emph{Phys. Lett. B}
  {\bfseries 749} (2015) 187}
  [\href{https://arxiv.org/abs/1504.03511}{{\ttfamily 1504.03511}}].

\bibitem{CMS:2013zfg}
{\scshape CMS} collaboration, \emph{{Measurement of the Differential and
  Double-Differential Drell-Yan Cross Sections in Proton-Proton Collisions at
  $\sqrt{s} =$ 7 TeV}},
  \href{https://doi.org/10.1007/JHEP12(2013)030}{\emph{JHEP} {\bfseries 12}
  (2013) 030} [\href{https://arxiv.org/abs/1310.7291}{{\ttfamily 1310.7291}}].

\bibitem{LHCb:2015okr}
{\scshape LHCb} collaboration, \emph{{Measurement of the forward $Z$ boson
  production cross-section in $pp$ collisions at $\sqrt{s}=7$ TeV}},
  \href{https://doi.org/10.1007/JHEP08(2015)039}{\emph{JHEP} {\bfseries 08}
  (2015) 039} [\href{https://arxiv.org/abs/1505.07024}{{\ttfamily
  1505.07024}}].

\bibitem{LHCb:2015mad}
{\scshape LHCb} collaboration, \emph{{Measurement of forward W and Z boson
  production in $pp$ collisions at $ \sqrt{s}=8 $ TeV}},
  \href{https://doi.org/10.1007/JHEP01(2016)155}{\emph{JHEP} {\bfseries 01}
  (2016) 155} [\href{https://arxiv.org/abs/1511.08039}{{\ttfamily
  1511.08039}}].

\bibitem{Accardi:2016qay}
A.~Accardi, L.T.~Brady, W.~Melnitchouk, J.F.~Owens and N.~Sato,
  \emph{{Constraints on large-$x$ parton distributions from new weak boson
  production and deep-inelastic scattering data}},
  \href{https://doi.org/10.1103/PhysRevD.93.114017}{\emph{Phys. Rev. D}
  {\bfseries 93} (2016) 114017}
  [\href{https://arxiv.org/abs/1602.03154}{{\ttfamily 1602.03154}}].

\bibitem{Accardi:2023gyr}
A.~Accardi, X.~Jing, J.F.~Owens and S.~Park, \emph{{Light quark and antiquark
  constraints from new electroweak data}},
  \href{https://doi.org/10.1103/PhysRevD.107.113005}{\emph{Phys. Rev. D}
  {\bfseries 107} (2023) 113005}
  [\href{https://arxiv.org/abs/2303.11509}{{\ttfamily 2303.11509}}].

\bibitem{Haario:2001}
H.~Haario, E.~Saksman and J.~Tamminen, \emph{An adaptive metropolis algorithm},
  {\emph{Bernoulli} {\bfseries 7} (2001) 223}.

\bibitem{Metropolis:1953}
N.~Metropolis, A.W.~Rosenbluth, M.N.~Rosenbluth, A.H.~Teller and E.~Teller,
  \emph{{Equation of State Calculations by Fast Computing Machines}},
  \href{https://doi.org/10.1063/1.1699114}{\emph{The Journal of Chemical
  Physics} {\bfseries 21} (2004) 1087}.

\bibitem{Hastings1970}
W.K.~Hastings, \emph{Monte carlo sampling methods using markov chains and their
  applications}, {\emph{Biometrika} {\bfseries 57} (1970) 97}.

\bibitem{HandbookMCMC:2011}
S.~Brooks, A.~Gelman, G.L.~Jones and X.-L.~Meng, \emph{Handbook of Markov Chain
  Monte Carlo}, Chapman \& Hall/CRC Handbooks of Modern Statistical Methods,
  CRC Press (2011).

\bibitem{Geyer:1992pmcmc}
C.J.~Geyer, \emph{{Practical Markov Chain Monte Carlo}},
  \href{https://doi.org/10.1214/ss/1177011137}{\emph{Statistical Science}
  {\bfseries 7} (1992) 473 }.

\bibitem{Link:2012thinning}
W.A.~Link and M.J.~Eaton, \emph{On thinning of chains in mcmc}, {\emph{Methods
  in ecology and evolution} {\bfseries 3} (2012) 112}.

\bibitem{Wolff2004}
U.~Wolff, \emph{{Monte Carlo errors with less errors}},
  \href{https://doi.org/10.1016/S0010-4655(03)00467-3}{\emph{Computer Physics
  Communications} {\bfseries 156} (2004) 143}.

\bibitem{Putze:2008ps}
A.~Putze, L.~Derome, D.~Maurin, L.~Perotto and R.~Taillet, \emph{{A Markov
  Chain Monte Carlo for Galactic Cosmic Ray physics: I. Method and results for
  the Leaky Box Model}},
  \href{https://doi.org/10.1051/0004-6361/200810824}{\emph{Astron. Astrophys.}
  {\bfseries 497} (2009) 991}
  [\href{https://arxiv.org/abs/0808.2437}{{\ttfamily 0808.2437}}].

\bibitem{Pumplin:2001ct}
J.~Pumplin, D.~Stump, R.~Brock, D.~Casey, J.~Huston, J.~Kalk et~al.,
  \emph{{Uncertainties of predictions from parton distribution functions. 2.
  The Hessian method}},
  \href{https://doi.org/10.1103/PhysRevD.65.014013}{\emph{Phys. Rev. D}
  {\bfseries 65} (2001) 014013}
  [\href{https://arxiv.org/abs/hep-ph/0101032}{{\ttfamily hep-ph/0101032}}].

\end{thebibliography}\endgroup
